\documentclass[%
 aip,
 sd,%
 amsmath,amssymb,
 reprint,%
]{revtex4-1}

\usepackage{graphicx}
\usepackage{dcolumn}
\usepackage{bm}

\begin{document}

\preprint{AIP/123-QED}

\title[Extracting physical chemistry from mechanics: a new approach...]{Extracting physical chemistry from mechanics: a new approach to investigate DNA interactions with drugs and proteins in single molecule experiments}

\author{M. S. Rocha}
 \email{marcios.rocha@ufv.br, mrocha.ufv@gmail.com}
\affiliation{ Laborat\'orio de F\'isica Biol\'ogica, Departamento de
F\'\i sica, Universidade Federal de Vi\c{c}osa. Av. P. H. Rolfs s/n,
CEP 36570-900, Vi\c{c}osa, MG, Brazil.
}%

\date{\today}

\begin{abstract}
In this review we focus on the idea of establishing connections
between the mechanical properties of DNA-ligand complexes and the
physical chemistry of DNA-ligand interactions. This type of
connection is interesting because it opens the possibility of
performing a robust characterization of such interactions by using
only one experimental technique: single molecule stretching.
Furthermore, it also opens new possibilities in comparing results
obtained by very different approaches, in special when comparing
single molecule techniques to ensemble-averaging techniques. We
start the manuscript reviewing important concepts of the DNA
mechanics, from the basic mechanical properties to the Worm-Like
Chain model. Next we review the basic concepts of the physical
chemistry of DNA-ligand interactions, revisiting the most important
models used to analyze the binding data and discussing their binding
isotherms. Then, we discuss the basic features of the single
molecule techniques most used to stretch the DNA-ligand complexes
and to obtain force $\times$ extension data, from which the
mechanical properties of the complexes can be determined. We also
discuss the characteristics of the main types of interactions that
can occur between DNA and ligands, from covalent binding to simple
electrostatic driven interactions. Finally, we present a historical
survey on the attempts to connect mechanics to physical chemistry
for DNA-ligand systems, emphasizing a recently developed fitting
approach useful to connect the persistence length of the DNA-ligand
complexes to the physicochemical properties of the interaction. Such
approach in principle can be used for any type of ligand, from drugs
to proteins, even if multiple binding modes are present.

\end{abstract}

\maketitle

\section{Introduction}

The DNA molecule is the biological polymer related to some of the
most important vital processes, from the storage and transmission of
genetic information to the translation of proteins. Its primary
structure is usually described as two parallel strands with a
peculiar chemical structure based in complementary base-pairs,
allowing the replication of the molecule in an unmistakable way
\cite{Watson2, Alberts}. The two DNA strands are arranged forming a
double-helix structure that sets important properties to the
molecule such as a well-defined negative charge density and a
bending stiffness which places DNA in the class of semi-flexible
polymers \cite{Bustamante, Marko, Odjik, Wang}.

Since it stores the genetic information of an organism, the DNA
molecule may be very long in some cases. In fact, the human genome
has approximately 3 billion base pairs, corresponding to a linear
contour length of the order of 1 meter. If a DNA molecule with this
length is placed disperse in a water-based solution, its radius of
gyration will be of the order of 100 $\mu$m \cite{deGenes}. How can
a molecule with this size be stored in the nucleus of a cell, which
has typical dimensions on the order of a few micrometers
\cite{Alberts}? The answer lies, at least partially, in the
mechanical properties of the DNA molecule, which must be unique to
allow such a condensation.  \textit{In vivo}, this process usually
occurs mediated by the interaction of the DNA molecule with ligands,
especially (but not exclusively) histone proteins. Furthermore, from
molecular biology it is known that other important intracellular
processes such as cell division and protein binding also depend on
the DNA topology, which in turn, depends on the mechanical
properties of the DNA molecule \cite{Bates, Vologodskiibook}. DNA
topology can be strategically changed during these processes by the
action of enzymes such as the topoisomerases, allowing their
occurrence efficiently \cite{Vologodskiibook, Vologodskii}.

Like the proteins and enzymes exemplified above, many drugs are
capable to interact with DNA, modifying its mechanical properties
with biological implications \textit{in vivo}. Cancer chemotherapy,
for instance, is a field in which the details about DNA interactions
with drugs are important. In fact, some classes of drugs such as the
anthracyclines and the platinum-based compounds exhibit a strong
affinity to interact with the DNA of cancer cells. When these drugs
bind to DNA they can inhibit the replication process, thus stopping the
tumor growth \cite{Hurley, Scripture}. On the other hand, gene
therapy is another field of medical sciences in which this kind of
knowledge is also important \cite{Li, Sheridan}. In these therapies,
DNA molecules are usually transported from outside to inside living
cells in order to replace defective genes, thus correcting cell
malfunctions. One approach to accomplish this transport easily, for
example, is condensing the DNA molecule by using cationic ligands
\cite{Hansma, Sullivan}.

In summary, all the examples discussed above show the importance in
studying and understanding the details behind DNA interactions with
ligands. In fact, many researchers of varied areas such as physics,
chemistry, biology, medicine, pharmacy, engineering, etc have paid
attention to this topic along the past 20 years, with a fast
increase of the number of publications and citations
\cite{publications}.

In this manuscript we review important topics of the field
``DNA-ligand interactions'', emphasizing in how one can connect the
changes of the mechanical properties of the DNA induced by the
binding ligand to the physicochemical information of such
interaction. This type of connection is interesting because it
allows one to perform a robust characterization of the interaction
both from the mechanical and physicochemical point of view by using
only one experimental technique: single molecule stretching
experiments. To discuss such connection, firstly in Section 2 we
discuss the basic DNA mechanics, revisiting the main concepts and
approaches used in the field. In particular, we revisit the
Worm-Like Chain (WLC) model, the standard one used to describe bare
DNA mechanics and to investigate the changes of the DNA mechanical
properties when interacting with a binding ligand. Then, in Section
3 we discuss the physical chemistry of DNA-ligand interactions,
emphasizing the chemical equilibrium states which can usually be
described by a binding isotherm. We revisit the most important
models used in the field, discussing their strong points and
limitations. In Section 4 we discuss briefly the experimental
techniques most used to perform single molecule experiments,
emphasizing the key features of each one. In Section 5 we present
and discuss the main types of interactions that occur between DNA
and ligands: intercalation, covalent binding, electrostatic driven
interactions and groove binding. Finally, in Section 6 we present
and discuss the main topic of this review: the approaches on how one
can connect mechanics to physical chemistry. The final conclusions
are presented in Section 7.

\section{DNA mechanics}

During the last decades, DNA mechanics has become a very well
studied topic especially due to the advent of single molecule
techniques. Such techniques allow one to manipulate and stretch
individual DNA molecules, giving access to mechanical information
contained in the ``force $\times$ extension'' curves. Before single
molecule techniques, such type of information was somewhat
difficult to be accessed by ensemble-averaging techniques.

From the mid-90s some theoretical models were formulated in order to
explain the mechanical behavior of DNA molecules. In particular,
most of these models attempt to give a theoretical expression for
the ``force $\times$ extension'' curve based on key mechanical
parameters such as the linear contour length of the polymer chain
and the DNA bending stiffness, which can be conveniently represented
by its persistence length.

The contour length is the most basic mechanical property of a
polymer chain: it is simply the length of the chain measured along
its contour, which is proportional to the number of monomers. The
persistence length, otherwise, is the correlation length of the
polymer chain and thus gives information about the bending stiffness
of the polymer. In the case of DNA molecule, the persistence length
has basically two components: the intrinsic and the electrostatic
one. The first component is related to the bending rigidity due to
the molecule composition itself, while the second one is due to the
negative charge distribution along the double-helix \cite{Odijk,
Manning, Baumann2, Wenner}. Since these two components are usually
present in most relevant situations, the models in general represent
the persistence length by its effective value, which takes into
account the two contributions. Following most authors, in this
manuscript we will call the effective persistence length only by
persistence length. In water-based solutions under nearly
physiological conditions (pH = 7.4, [NaCl] = 150 mM), a disperse
bare DNA molecule is classified as a semi-flexible (or semi-rigid)
polymer due to its intermediate value of the bending stiffness,
which corresponds to a persistence length $A$ $\simeq$ 50 nm
\cite{Wang, Strick, Smith, Boal}.

In the following section we present the most relevant model used for
studying the mechanics of DNA molecule: the Worm-Like Chain (WLC)
model. In this manuscript we do not intend to review other
mechanical models or present an historical survey on this specific
subject, since today the WLC model is recognized as the standard one
to study DNA mechanics.

\subsection{Worm-Like Chain Model (WLC)}\label{WLC}

The Worm-Like Chain (WLC) is a model derived from polymer physics,
and has become in the past years the standard one in analyzing DNA
stretching experiments. To introduce this model, let us firstly
assume that the polymer itself is a chain formed by rigid rods with
lengths $b$, connected by freely-rotating vertices. Let us call
$\theta_i$ the angle between the rods $i$ and $i$ + 1. The WLC model
is then defined by assigning an harmonic bending energy function to
the angle formed between the two rods \cite{Schellman, Wiggins1, Wiggins2,
Marko, Boal, Vologodskii2},

\begin{equation}
E(\theta_i) = \frac{\kappa}{2b}\theta_i^2\label{WLC1},
\end{equation}
taking the continuum limit with $b \rightarrow$ 0. The constant
$\kappa$ is the effective elastic bending stiffness of the chain.

In the continuum limit, Eq. \ref{WLC1} can be used to write the
total bending energy of the chain \cite{Marko, Boal},

\begin{equation}
E = \frac{\kappa}{2}\int_0^L|C|^2ds\label{WLC2},
\end{equation}
where $C$ is the local curvature at each point, $ds$ is a length
element along the polymer, and $L$ is the contour length of the
polymer chain.

The parameter $\kappa$ is directly related to the polymer
persistence length $A$ by

\begin{equation}
A = \frac{\kappa}{k_BT}\label{WLC3},
\end{equation}
where $k_B$ is Boltzmann's constant and $T$ is the absolute
temperature.

Equation \ref{WLC2} can be used to deduce the behavior of the force
as a function of the polymer extension as one stretches it. This
analysis can be performed numerically or analytically using
appropriate approximations \cite{Boal}. In 1995, Marko and Siggia
solved the model analytically, obtaining an approximate expression
for the force as a function of the polymer extension which has
become the most used to analyze DNA stretching experiments in the
entropic low-force regime ($F \leq$ 5 pN) \cite{Marko, Smith}. Their
result is

\begin{equation}
F = \frac{k_BT}{A}\left[\frac{z}{L} + \frac{1}{4\left(1 -
\frac{z}{L}\right)^2} - \frac{1}{4}\right] \label{markosiggia},
\end{equation}
where $F$ is the force and $z$ is the end-to-end distance
(extension) of the DNA molecule.

Despite its renowned utility, this expression is still an
approximation, diverging at $z$ = $L$. Moreover, Eq.
\ref{markosiggia} describes well only the entropic regime of the
polymer, which is valid for stretching forces typically below $\sim$
5 pN. In this regime the applied forces are sufficiently small such
that they can change only the polymer conformation in solution,
\textit{i. e.}, its entropy.

Also in 1995, Odjik proposed a different approach that accounts for
higher forces, in which enthalpic effects start to became relevant
for the polymer mechanics \cite{Odjik}. The enthalpic regime is
defined as the regime in which the stretching forces became large
enough to distort the DNA primary structure and eventually to break
chemical bonds. Such effect can be accounted by introducing an
enthalpic mechanical parameter to describe the polymer deformation:
the stretch modulus $S$. The analytical expression proposed by Odjik
reads \cite{Odjik}

\begin{equation}
z = L\left[1 - \frac{1}{2}\sqrt{\frac{k_BT}{AF}} + \frac{F}{S}
\right]\label{WLC_Odjik}.
\end{equation}

Observe that the stretch modulus $S$ has units of force. Taking the
limit $S \rightarrow \infty$ and inverting the above equation
(isolating $F$), we found an equation similar to the Marko-Siggia
expression (Eq. \ref{markosiggia}) if $z \sim L$, \textit{i. e.},
neglecting very small forces. Thus, observe that a polymer in the
entropic regime can be interpreted as a polymer that has a stretch
modulus $S$ very large, \textit{i. e.}, that resists deformations on
its chemical structure.

In 1999, Bouchiat \textit{et al.} proposed another solution of the
WLC model in the entropic regime. Their approach consists in adding
six terms to Eq. \ref{markosiggia} in order to improve its accuracy
\cite{Bouchiat}. These terms were determined by comparing the
results predicted by Eq. \ref{markosiggia} to results from an exact
numerical solution of the WLC model \cite{Bouchiat}, which was done
perturbatively. The resulting expression reads

\begin{equation}
F = \frac{k_BT}{A}\left[\frac{z}{L} + \frac{1}{4\left(1 -
\frac{z}{L}\right)^2} - \frac{1}{4} + \sum_{i = 2}^7
a_i\left(\frac{z}{L}\right)^i\right] \label{expbouchiat},
\end{equation}
where the $a_i$'s are constants numerically determined.

In addition to the models discussed above, important contributions
to the elucidation of many peculiarities of DNA mechanics were given
by the groups of A. Vologodskii, M. D. Frank-Kamenetskii, H. E.
Gaub, M. C. Williams, V. Croquette, F. Ritort, C. Bustamante and
others, especially concerning the bending of small DNA fragments,
strong bending and fluctuations in the double-helix, dependence of
DNA rigidity on the temperature and base sequence, DNA twist,
overstretching transition, DNA hairpins, etc. \cite{Vologodskii2,
Vologodskii3, Kamenetskii, Geggier, Geggier2, Kamenetskii3,
Kamenetskii2, Clausen, Rief, Wenner, Williams2009, Strick, Strick2,
Charvin, Mossa, Manosas, Huguet, Alemany, Smith2, Bustamante2000,
Baumann, Bustamante2003, Gore, Bryant, Bryant2012}.

\section{Physical chemistry of DNA-ligand interactions}\label{physchem}

The study of the physical chemistry of DNA-ligand interactions
consists in two different sub-fields: the chemical equilibrium of
the interaction and the kinetics of the interaction. Consider the
system of interest (DNA + ligand molecules in solution) as composed
by two different partitions where the ligand molecules can stay: the
DNA (bound ligand molecules) and the solution (free ligand
molecules). The chemical equilibrium is achieved when the average
number of molecules in the partitions remains constant in time. The
kinetics of the interaction, otherwise, describes the changes that
occur between the initial incubation and the final equilibrium
state.

In this manuscript we emphasize the physical chemistry of the
chemical equilibrium, since the equilibrium states can be
represented by a binding isotherm that can be linked to the changes
of the mechanical properties of DNA-ligand complexes. Below we
discuss the most relevant models that attempt to describe the
chemical equilibrium of DNA-ligand interactions. Some studies on the
kinetics of such interactions were performed by the groups of M. C.
Williams, D. Anselmetti, D. M. Crothers and others \cite{Pant,
Paramanathan, Paramanathan2, Cruceanu, Sischka, Kleimann, Slutsky,
Phillips, Chaires2}.

\subsection{The general problem}\label{general}

Consider two molecules A and B associating in solution to result in
a molecule C. This mechanism can be represented by the chemical
reaction

\begin{equation}
A + B \mathop{\rightleftharpoons}^{K_i}_{K_d} C, \label{reacao1}
\end{equation}
where $K_i$ and $K_d$ are, respectively, the equilibrium intrinsic
binding constants of association and dissociation. They are also
known as thermodynamic constants or macroscopic constants. Observe
that $K_i$ represents the association reaction, where the reagents A
and B associate to result in the compound C, while $K_d$ represents
the dissociation reaction, \textit{i. e.}, the reverse reaction in
which C dissociate in the original reagents A and B.

These constants are defined in term of the molar concentrations of
the involved substances,

\begin{equation}
K_i = \frac{[C]}{[A][B]}, \label{defKi}
\end{equation}
and

\begin{equation}
K_d = \frac{[A][B]}{[C]} = K_i^{-1}. \label{defKd}
\end{equation}

Note that in these last two equations, [$X$] is the molar
concentration (1 M = 1 mol/liter) of the compound $X$. Also observe
that $K_i$ has units of M$^{-1}$, while $K_d$ has units of M.

\subsection{Scatchard model}\label{scat}

This is the simplest model that describes the chemical equilibrium
of the DNA molecule with ligands in solution. Let us firstly adapt
the previous notation for the specific case of DNA-ligand
interactions. Call [$A$] $\equiv$ $C_f$ the concentration of free
ligands solution and [$C$] $\equiv$ $C_b$ the concentration of
ligands bound to DNA (result of the reaction). Suppose firstly that
each ligand molecule occupies only one base pair of the DNA when
bound. Consequently, the concentration of free linkable sites in the
DNA molecule can be written as [$B$] $\equiv$ $C_{bp}$ - $C_b$,
where $C_{bp}$ is the concentration of DNA base pairs, which is a
constant.

Substituting these definitions in Eq. \ref{defKi}, one has

\begin{equation}
K_i = \frac{C_b}{C_f(C_{bp} - C_b)}.\label{Scat1}
\end{equation}

Now we introduce the bound ligand fraction $r$,

\begin{equation}
r = \frac{C_b}{C_{bp}},\label{defr}
\end{equation}
such that Eq. \ref{Scat1} can be rewritten as

\begin{equation}
r = \frac{K_iC_f}{1 + K_iC_f},\label{Scatfinal}
\end{equation}
which is known as the Scatchard binding isotherm, proposed
originally in 1949 \cite{Scatchard}.

Despite its didactic utility, the Scatchard binding isotherm has two
important simplifications: (a) It is valid only for very small
ligand molecules which occupy only one DNA base-pair when bound,
which is not the case for most ligand molecules. (b) It supposes
that previous bound ligand molecules do not interfere in the binding
mechanism of the subsequent ones, \textit{i. e.}, the interaction is
non-cooperative.

The first simplification can be bypassed by introducing the
parameter $r_{max}$, the bound ligand fraction at saturation,
\textit{i. e.} the maximum value of the bound ligand fraction $r$.
Observe that the inverse of $r_{max}$ is the mean number of base
pairs occupied by each bound ligand molecule $N$ = 1/$r_{max}$. The
corrected binding isotherm then reads

\begin{equation}
r = \frac{r_{max}K_iC_f}{1 + K_iC_f},\label{Scatfinalrmax}
\end{equation}

\subsection{Hill model}\label{hill}

The Hill binding isotherm was originally proposed by A. V. Hill in
1910 to describe the binding of oxygen to hemoglobin inside red
blood cells \cite{Hill}.

Basically the model introduces the Hill exponent $n$, a
cooperativity parameter which is a lower bound for the number of
cooperating ligand molecules involved in the reaction \cite{Siman,
Cesconetto}. The binding isotherm reads

\begin{equation}
r = \frac{r_{max}(K_iC_f)^n}{1 + (K_iC_f)^n}.\label{Hillfinal}
\end{equation}

The apparent binding association constant of the reaction is defined
as $K_A$ = $K_i^n$. Observe that if $n >$ 1, the interaction is
positively cooperative, \textit{i. e.}, a bound ligand molecule
increases the apparent affinity of DNA for subsequent ligand
binding. If $n <$ 1, otherwise, the interaction is negatively
cooperative and a bound ligand molecule decreases the apparent
affinity of DNA for subsequent ligand binding. If $n$ = 1, the
interaction is non-cooperative and the affinity is independent of
the number of previously bound ligand molecules.

The Hill binding isotherm has achieved a particular success to
describe positively cooperative ``none-or-all'' processes ($n >$ 1),
in which the cooperating ligand molecules bound practically
simultaneously to the bound site forming a bound cluster
\cite{Siman, Silva}. On the other hand, when $n$ = 1 the Hill
isotherm reduces to the Scatchard one and is therefore able to
describe individual binding of ligand molecules \cite{Cesconetto}.
Finally, to the best of our knowledge there is no report in the
literature of a negatively cooperative DNA-ligand interaction
described by a Hill binding isotherm.

\subsection{Neighbor exclusion model (NEM)}\label{NEM}

This model was proposed in 1974 by McGhee and von Hippel with the
purpose of analyzing in detail the neighbor exclusion effects due to
large ligand molecules that occupy more than one DNA base-pair
\cite{McGhee, RochaBioPol}. The authors have accounted for the
ligand size by introducing the exclusion parameter $N$, the number
of base-pairs that a ligand molecule effectively occupies when
binding to DNA. This parameter was cited earlier in connection to
the saturated bound ligand fraction, $N$ = 1/$r_{max}$.

The model has a non-cooperative and a cooperative version, but the
last one has been used only in a few works \cite{McCauley2013,
Murugesapillai} to analyze experimental data because the binding
isotherm is somewhat intricate.

The non-cooperative binding isotherm reads

\begin{equation}
\frac{r}{C_f} = K_i(1 - Nr)\left[\frac{1 - Nr}{1 - (N -
1)r}\right]^{N - 1} \label{nemfinal},
\end{equation}
and the cooperative binding isotherm reads

\begin{eqnarray}
\frac{r}{C_f} = K_i(1 - Nr)\left[\frac{(2\omega - 1)(1 - Nr) +r -
R}{2(\omega - 1)(1- Nr)}\right]^{N - 1}\nonumber\\
\times\left[\frac{1 - (N + 1)r + R}{2(1 - Nr)}\right]^2
\label{nemcoopfin},
\end{eqnarray}
with

\begin{equation}
R = \sqrt{[1 - (N + 1)r]^2 + 4\omega r(1 - Nr)} \label{nemcoopR}.
\end{equation}

Here $\omega$ is the cooperativity parameter. For $\omega$ smaller,
equal, or larger than unity, one has negative, non-cooperative, or
positive cooperativity, respectively.

The major advantage of this model is to treat in more detail the
effects related to the ligand size. This feature is particularly
important in the analysis of DNA interactions with intercalators, a
class of ligands in which neighbor-exclusion effects is extremely
important \cite{LDWilliams92, Brana, Goftar}. In fact, NEM has
become in the past years the standard binding isotherm used to
analyze DNA interactions with intercalators \cite{Chaires, Gaugain,
RochaJCP2, Reis, Crisafuli2014}.

\section{Single molecule experimental methods}

In this section we briefly discuss the single molecule experimental
techniques commonly used to measure the mechanical properties of the
DNA-ligand complexes.

The main advantage of single molecule techniques is the possibility
to study a particular DNA molecule free from the influence of other
molecules in the sample. Single molecule stretching experiments such
as those performed with optical or magnetic tweezers usually give
insights on the global (long length scale) mechanical properties of
individual DNA molecules. In fact, mechanical parameters such as the
persistence and contour lengths and the stretch modulus can be
extracted by analyzing the force $\times$ extension curves of the
complexes, which can be obtained in single molecule approaches.

Useful reviews which discuss and compare single molecule techniques
can be found in the literature \cite{Ritort2, Neuman2, Neuman3}.

\subsection{Optical tweezers}

Since the seminal works of Ashkin and collaborators \cite{Ashkin1,
Ashkin2}, optical trapping and manipulation have found various
applications in many areas of science such as physics, biology and
chemistry. Today, the most common optical tweezers are mounted by
focusing a laser beam with a microscope objective of large numerical
aperture. This apparatus can trap small dielectric objects near the
lens focus, being a powerful tool to manipulate beads, particles and
biological systems with typical sizes in the micrometer range
\cite{Ashkin4, Ashkin2}. The typical forces obtained with this
apparatus are between 0.1 - 400 piconewtons, which are in the range
of many biological forces such as the entropic and enthalpic forces
on biopolymers and molecular motors. For an introductory review
about the basic theory and features of optical tweezers, see ref.
\cite{RochaAJP2}. Other useful reviews on instrumentation and recent
advances on the technique can also be found in the literature
\cite{Moffitt, Grier, Svoboda, Neuman}.

To perform precise quantitative measurements with optical tweezers,
size-calibrated dielectric beads have become the standard objects to
be captured because of their perfect symmetry which facilitates trap
calibration and position detection. A dielectric bead trapped in an
optical tweezers is an overdamped Brownian harmonic oscillator, such
that the optical trap can be characterized by its trap stiffness
$\kappa$ which depends on the bead size and refractive index
\cite{RochaAJP2}.

In the last decades, optical tweezers have been largely used to
study the mechanical properties of DNA/RNA molecules and their
complexes formed with drugs or proteins. Useful reviews on this
subject can be found in the literature \cite{Heller, Chaurasiya,
McCauley, McCauley2}. Basically, the classic experiment consists in
attaching one end of the DNA molecule to a polystyrene or silica
bead and the other end to a substrate (a microscope coverslip or a
second bead attached to a micropipette, for example). The optical
tweezers is then used to trap the bead and so the DNA molecule can
be manipulated and stretched by moving the laser beam or the
microscope stage. The force as a function of extension can be
measured as one stretches the DNA molecule. To perform this task,
one needs to detect the bead position and to calibrate the tweezers
(determine the trap stiffness $\kappa$). There are many techniques
which can be used to perform this kind of measurement, such as
dynamic light scattering \cite{Shivashankar, Rocha}, back-focal
plane interferometry \cite{Allersma}, statistics of thermal
fluctuations \cite{Meiners}, simple videomicroscopy
\cite{CrisafuliIB, CrisafuliAPL, Cesconetto}, calibration using
hydrodynamic drag forces \cite{Sischka} or by using other types of
detectors \cite{McCauley}. For a recent review on measuring with
optical tweezers, see ref. \cite{Moffitt}.

\begin{figure}
\centering
\includegraphics[width=6.5cm]{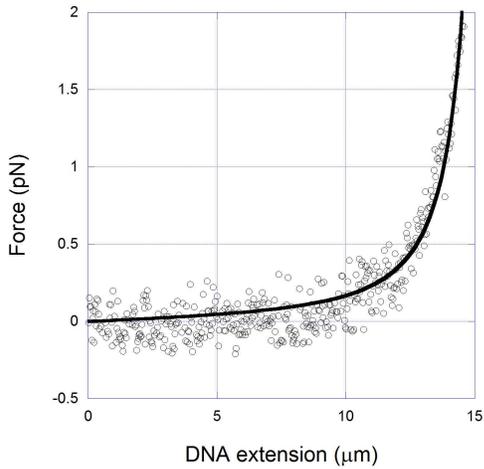}
\caption{Force $\times$ extension curve of a bare $\lambda$-DNA molecule in
the entropic regime. \textit{Circles}: experimental data obtained
with optical tweezers; \textit{Solid line}: a fitting to the
Marko-Siggia Worm-Like Chain (WLC) model (Eq. \ref{markosiggia}).
For this particular DNA molecule we have found from the fitting $A$
= (50 $\pm$ 2) nm and $L$ = (15.6 $\pm$ 0.1) $\mu$m.}
\label{forceDNA}
\end{figure}

Figure \ref{forceDNA} shows a typical force $\times$ extension curve
of a single bare $\lambda$-DNA molecule ($\sim$ 48,500 base-pairs)
obtained by performing a DNA stretching experiment in the entropic
regime with optical tweezers. The trap calibration and the bead
position detection were performed in this case by using
videomicroscopy \cite{CrisafuliIB}, and the solid line corresponds
to a fitting using the Marko-Siggia WLC model (Eq.
\ref{markosiggia}). From the fitting one can promptly determine the
persistence and contour lengths of the DNA molecule, obtaining for
this particular curve $A$ = (50 $\pm$ 2) nm and $L$ = (15.6 $\pm$
0.1) $\mu$m.

\subsection{Magnetic tweezers}

The idea behind magnetic tweezers is very similar to its optical
analogue, the main difference is that in this case the forces are
exerted by an external magnetic field applied around the sample.
Paramagnetic beads are used instead of dielectric ones in order to
be manipulated with the magnetic field. The typical forces obtained
are of the order of hundredths of piconewtons to hundreds of
piconewtons.

Basically, the force applied on the paramagnetic beads can be
written as

\begin{equation}
\overrightarrow{F} = -
\frac{1}{2}\overrightarrow{\nabla}(\overrightarrow{\mu}\cdot\overrightarrow{B})
\label{fordipmag},
\end{equation}
where $\overrightarrow{\mu}$ is the magnetic dipole moment induced
in the bead and $\overrightarrow{B}$ is the applied magnetic field.
Observe that for moderate magnetic fields one has
$\overrightarrow{\mu}$ $\propto$ $\overrightarrow{B}$ and the
resulting force is proportional to the gradient of the field
intensity.

Reviews on the basic and advanced features of magnetic tweezers can
be found in the literature \cite{Vlaminck, Kilinca, Lipfert2}.

An advantage of this technique in relation to optical tweezers is
its convenience to apply torques on the magnetic beads by rotating
the external magnetic field, which allows one to rotate DNA
molecules and therefore to study quantities such as the torsional
rigidity and the degree of supercoiling \cite{Gosse, Strick,
Strick2, Salerno, Bryant, Moroz, Celedon}. These quantities are also
mechanical properties important to some biological processes in
which the double-helix must be unwound, such as in DNA replication.
Another advantage of the magnetic tweezers is its convenience to
perform constant-force experiments, working as a force-clamp trap
(it is just a matter of choosing the adequate magnetic field - see
Eq. \ref{fordipmag}). Constant-force experiments can be performed
with optical tweezers only using non-conventional (and more
intricate) approaches such as by using a force-feedback electronics
or working in anharmonic regions of the optical potential
\cite{Nambiar, Greenleaf}. Among the disadvantages of using magnetic
tweezers, one can cite the hysteresis of the magnetic field and heat
generation around the sample if the field is produced by current
distributions, aside the more intricate calibration of the apparatus
and its restriction to applications with magnetic materials. A
recent work by Neuman and Nagy provides a detailed comparison
between optical and magnetic tweezers, and also atomic force
microscopy \cite{Neuman2}.

\subsection{Atomic force microscopy (AFM)}

Atomic force microscopy (AFM) is another important tool in
single-molecule studies of DNA-ligand interactions
\cite{BustaRivetti, Anselmetti2000}. In the last 20 years, a number
of different protocols have been developed in order to deposit DNA
molecules on a flat surface and to image them reliably and
reproducibly. Today, the standard surfaces used to deposit the DNA
molecules are mica substrates and less often silicon substrates,
because of their low rugosity. A number of buffer solutions
containing divalent cations (such as Mg$^{2+}$, Ni$^{2+}$,
Mn$^{2+}$, Co$^{2+}$ and Ca$^{2+}$) have been used to enhance the
DNA adsorption onto the substrate, which is otherwise poor.
Moreover, divalent cations also allow the polymer chain to
equilibrate on the flat 2D adsorbing surface, preventing chain
kinetic trapping, which must be avoided in order to study
equilibrium properties of the adsorbed DNA or DNA-ligand complex
\cite{BustaRivetti}. Once adsorbed, DNA molecules and DNA-ligand
complexes can be imaged using the AFM usually operating in the
tapping mode, which minimizes possible damages to the sample due to
the tip-surface interactions during the scanning. The images
obtained are topographical maps which associate a certain height to
each point on the sample. By analyzing these images, several DNA
statistical parameters such as the mean contour length, the
persistence length and bending angles can be estimated. As discussed
by Rivetti \textit{et al.} \cite{Rivetti}, this analysis can be
performed, for example, by measuring the mean-squared end-to-end
distance $<R^2>$ of the polymer. In fact, statistical mechanics of
polymers predicts that for 2D worm-like chains (which is the case of
deposited DNA molecules), $<R^2>$ is given by

\begin{equation}
<R^2> = 4AL\left[1 - \frac{2A}{L}\left(1 -
e^{-\frac{L}{2A}}\right)\right] \label{2Ddna}.
\end{equation}

By using this equation it is possible determine the persistence
length $A$ by measuring $<R^2>$ and the contour length $L$ for the
deposited DNA molecules.

On the other hand, the visualization of DNA condensates formed with
polycations with the AFM technique is nowadays a routine in many
laboratories. The morphology of these condensed DNA complexes seen
in the AFM images are as well as clear and well-defined as the
images produced by other kinds of microscopy techniques such as
electron microscopy (EM). The structure of DNA-protein complexes has
also been a target of a number of studies. Within the limits of the
AFM technique are, for instance, the visualization of sharp kinks
and cross-links introduced in the DNA molecule by histone-like
proteins, the structure of nucleosome particles, the determination
of protein binding-sites and more recently the determination of
protein association constants to DNA \cite{Yang}.

Finally, besides being a powerful tool for visualizing single
molecules, the AFM apparatus can also be used to perform force
spectroscopy in solution like optical or magnetic tweezers, allowing
one to determine the force $\times$ extension curves of the
DNA-ligand complexes \cite{Clausen, Strunz, Zlatanova}.

Specific reviews about the application of the AFM technique to
single molecule studies can be found in the literature
\cite{Zlatanova, Engel}.

As a final remark, with the improvement of fluorescent-based optical
technology along the last decades, fluorescence microscopy has also
became another important tool to visualize DNA structure,
conformation changes and interactions with ligands at single
molecule level \cite{Yoshikawa92, Yoshikawa, Yoshinaga, Kurtz,
Yoshikawa2010, Amitani, Mameren, King, CeZhang}. The technique can
be used as complementary to AFM, with the advantage that one does
not need to deposit the molecules in a substrate.

\section{DNA-ligand interactions}

DNA can interact with ligands in many different ways, from covalent
binding to simple electrostatic driven interactions. Here we
describe briefly the most relevant types of interactions, discussing
the main features of each one.

\subsection{Covalent ligands}

The covalent binding of drugs to DNA is usually irreversible and
completely inhibit DNA processes. The platinum-based compounds are
examples of drugs which can interact with DNA by covalent binding
\cite{Strothkamp, Cohen}. Cisplatin and its related compounds
carboplatin and oxaliplatin are antitumor platinum-based molecules
usually used in cancer chemotherapy. The action of these complexes
as anticancer drugs consists in damaging the DNA molecule with
adducts that form various types of crosslinks, which introduce
strong structural perturbations and impede DNA replication
\cite{Lee, Hou}. The clinical use of these complexes, however, is
limited due to their several side effects and the development of
drug resistance.

Another example of covalent binding is found in the interaction of
drugs from the class of furocoumarins (psoralen, angelicin, etc)
with DNA when one illuminates the complex with ultraviolet-A (UVA)
light \cite{Sinden}. Psoralen is a well-known drug used in the
treatment of skin diseases like psoriasis, vitiligo, and some other
kinds of dermatitis \cite{McNeely}. The most common therapy is
called PUVA (psoralen followed by UVA light), which consists in
taking a medicine containing psoralen and exposing the patient to
UVA light. The drug effectively increases the skin sensitivity to
UVA and the skin melanin level \cite{Coven, Tran}. It is well
established in the literature that when a DNA-psoralen complex is
illuminated with UVA light, the drug molecules absorb photons and
form covalent bonds preferentially with the thymines \cite{Sinden,
Spielmann}. When there is no illumination at the sample, however,
psoralen interacts with DNA by intercalative binding - see next
section. The effects of covalent binding on the mechanical
properties of the DNA-psoralen complexes were recently studied
\cite{Rocha, RochaAPLPso}. In particular, it was shown that the
contour and persistence lengths of the complexes depend on the
psoralen concentration and on the exposure time to UVA light
\cite{RochaAPLPso}.

\subsection{Intercalators}

Intercalative binding is one of the most common interactions between
DNA and ligands, and was firstly described by L. S. Lerman in 1961
\cite{Lerman1, Lerman2}. It is characterized by the insertion of a
flat aromatic molecule between two adjacent DNA base pairs. The
complex is thought to be stabilized by the stacking interactions
between the ligand and the DNA bases \cite{Lin}. Intercalators also
introduce strong structural perturbations on the double-helix
structure. To accommodate the intercalated molecules, there is an
increase in the DNA contour length, which is accompanied by an
unwinding of the double-helix by a certain angle per intercalated
molecule \cite{LDWilliams92, Brana, Goftar, Sischka, Chaires,
Fritzsche}. Daunomycin, doxorubicin and ethidium bromide (EtBr) are
classic examples of drugs which intercalate in the DNA molecule and
can modify its elasticity depending on the drug concentration.
Daunomycin and doxorubicin are anthracycline antibiotics used in the
treatment of various cancers such as some types of leukemias,
sarcomas, lymphomas, myelomas, neuroblastomas, as well as cancers in
the breast, head, ovary, pancreas, prostate, stomach, liver, lung
and others. They inhibit DNA replication and transcription when
intercalating, impeding cell duplication \cite{Chaires}. Ethidium
bromide (EtBr) is commonly used as a fluorescent stain for
identifying and visualizing nucleic acid bands in electrophoresis
and in other methods of nucleic acid separation. Other known
intercalators are the DNA fluorescent stains acridine orange,
methylene blue \cite{Nafisi} and diaminobenzidine \cite{Reis}. More
examples and specific reviews on the basic properties of
intercalators can be found in the literature \cite{LDWilliams92,
Brana, Goftar}.

Many aspects of the DNA-daunomycin interaction, such as kinetics,
self association and equilibrium binding were studied by J. B.
Chaires, D. M. Crothers and coworkers in the 80's \cite{Fritzsche,
Chaires, Chaires2, Chaires3}. On the other hand, the DNA-EtBr
interaction was characterized in various aspects by many authors,
but even today one can found somewhat contradictory results about
the mechanical behavior of such complexes \cite{Smith, Tessmer,
RochaJCP2, Sischka, Cassina, Lipfert, Crisafuli2014}, and also for
different complexes formed between DNA and other intercalators,
especially when comparing results obtained  from different
experimental techniques \cite{Tessmer, RochaJCP2, Sischka, Cassina,
Lipfert, Matsuzawa, Quake, Berge, Yoshikawa92, Kaji, Reis,
Crisafuli2014}.

Recently our group has studied in detail the changes in the
persistence and contour lengths of DNA complexes formed with various
intercalating molecules \cite{Rocha, RochaJCP2, RochaAPLPso, Reis,
Crisafuli2014}, by using optical tweezers in a very low force regime
($F <$ 2 pN). We reported an abrupt structural transition in the
persistence length due to drug intercalation, which is probably
related to a partial denaturation of the DNA molecule due to the
pulling force used to stretch the complexes \cite{RochaPB,
RochaAPLPso, Reis, Crisafuli2014}. The contour length, otherwise,
does not present such a transition, increasing monotonically with
drug concentration until saturation.

\subsection{Electrostatic driven interactions}

Since the DNA molecule has a high negative charge density in aqueous
solution due to its phosphates (2 elementary charges per each 3.4
{\AA} along the DNA axis), it strongly interacts with itself
(\textit{i. e.}, different DNA segments strongly repel each other
thus promoting the chain swell) as well as with positively charged
ligands such as ions and macro-ions, especially multivalent cations
\cite{Bloomfield2}.

DNA condensation due to multivalent cations is a classic example
which shows the importance of electrostatic driven interactions in
DNA solutions \cite{Bloomfield2, Bloomfield, Arscott, Plum}. In this
process the multivalent cations binds along the DNA double-helix,
and the strong positional correlations between them start to play a
role and promotes a coil-globule transition: the DNA molecule folds
onto itself \cite{Gronbech, Winkler, Khan} with a high increase in
the local DNA segment density at the level of both mono-molecular
collapse or in a multi-molecular aggregation.

Some models concerning electrostatic interactions between DNA and
ligands were proposed along the last decades \cite{Rouzina3,
Podesta, Manning}. In fact, there are different hypotheses to
explain the DNA bending mechanism by multivalent cations, including
a purely electrostatic model by Rouzina and Bloomfield
\cite{Rouzina3} and an asymmetrical phosphate neutralization model
by Manning \cite{Manning}. According to Rouzina and Bloomfield, a
multivalent cation binds to the entrance of the DNA major groove,
between the two phosphate strands, electrostatically repelling
sodium counterions from the neighboring phosphates. The unscreened
phosphates on both strands are strongly attracted to the
groove-bound cation. This binding leads to groove closure,
accompanied by DNA bending towards the cationic ligand
\cite{Rouzina3}. Differently, Manning proposes that the stable
double-helix structure of DNA represents an equilibrium between
stretching forces (caused by interphosphates repulsion) and
compressive forces (caused by attractive interaction between
nucleotides). This analysis suggests that significant local
interphosphate stretching forces balance compressive forces within
DNA and that these stretching forces can drive DNA deformation when
phosphates charge are locally neutralized.

These two approaches predict a reduction of the persistence length
as the concentration of bound cations increases. In fact, the model
proposed by Rouzina and Bloomfield predicts that the effective
persistence length $A_E$ of the DNA-ligand complex is given by

\begin{equation}
\frac{1}{A_E} = \frac{1}{A_1} + \frac{Nr}{A_2} \label{modRouz},
\end{equation}
where $A_1$ is the bare DNA persistence length (when no ligands are
bound $r$ = 0) and 1/$A_1$ + 1/$A_2$ is the inverse persistence
length of a DNA saturated with ligands (which occurs when $r$ =
$r_{max}$ = 1/$N$). Observe that here $N$ is the exclusion parameter
of the ligand and $r$ = $C_b$/$C_{bp}$ is the ratio between the
bound ligand concentration and the DNA base-pair concentration, as
introduced in Section III.

The model developed by Manning, on the other hand, predicts that the
effective persistence length $A_E$ of the DNA-ligand complex (charge
neutralized DNA) is related to the original persistence length $A_0$
(fully charged DNA) by the equation \cite{Podesta, Manning}

\begin{equation}
A_E = \frac{2}{\pi R^2} \left[\frac{\beta A_0}{2(\xi - 1) -
\ln(\kappa b)}\right]^{3/2} \label{modMann},
\end{equation}
where $R$ is the radius of the double helix, $\beta$ is the Bjerrum
length (distance between two unit charges in pure solvent - no other
ions - at which the electrostatic energy is $k_BT$), 1/$\kappa$ is a
measure of the extent of the ion cloud around the object, $b$ is the
average axial distance between phosphates (0.17 nm) and $\xi$ =
$\beta$/$b$ is a measure of the axial charge density of the DNA
\cite{Podesta}.

This model predicts, for example, that for 30$\%$ of neutralized
charge, the effective persistence length is $A_E$ = 33.2 nm. For
60$\%$ of neutralized charge, $A_E$ = 11.1 nm and for 100$\%$ of
neutralized charge, $A_E$ = 7 nm \cite{Podesta}.

\subsection{Major and minor groove ligands}

Most drugs that interact electrostatically with DNA usually exhibits
a preference to the major or minor groove floor of the double-helix.
Many minor groove ligands are known by their antitumor and
antibiotic functions. This kind of interaction is usually
characterized by a combination of electrostatic, van der Waals and
hydrogen bonds. Examples of minor groove ligands are the anticancer
compound distamycin A, the antibiotics netropsin and berenil, and
the fluorescent stain DAPI. These drugs usually form reversible
complexes with DNA, preferentially binding at AT base pairs
sequences. They also induce elasticity changes on the DNA molecule,
stabilizing the double-helix structure \cite{Sischka}. An extensive
review on the DNA minor groove complexes can be found in ref.
\cite{Geierstanger}.

On the other hand, major groove binding is also a kind of
interaction usually characterized by electrostatic binding
\cite{Sischka}. $\alpha$-Helical (Ac-(Leu-Ala-Arg-Leu)3-NH linker)
is a peptide which interacts with DNA via major groove binding
\cite{Sischka}. Other known examples are the intercalator and major
groove ligand YO \cite{Eckel}, the bis-intercalator and major groove
ligands YOYO and ditercalinium \cite{Eckel, Berge, Hamilton} and the
anticancer drug neocarzinostatin \cite{Hamilton}. More examples and
a discussion on the main characteristics of major groove binding
ligand can be found in a recent review \cite{Hamilton}.

\subsection{Ligands with multiple binding modes}

There are many ligands which can interact to DNA by different
binding modes, depending on factors such as the properties of the
surrounding buffer solution, the DNA base-pair sequence, external
conditions such as sample illumination, etc. Some examples were
already cited in the last sections. In some cases the ligand has
distinct portions which interact to DNA by different modes. In other
cases there is only a single binding mode for the entire ligand
molecule, which can be changed upon determined conditions.

Bis-intercalators like YOYO and ditercalinium, for instance, are
molecules which have two intercalating portions linked by another
chemical structure which sometimes may interact with the DNA grooves
\cite{Eckel, Berge, Hamilton}. Actinomycin D is another example of a
drug with distinct portions that interact to DNA by different modes,
in this case including minor groove binding and intercalation
\cite{Patel, Muller, Sobell, Takusagawa, Cesconetto}.

Psoralen is an example of a drug which the binding mode depends on
an external condition (sample illumination). As explained before,
the drug initially intercalates in DNA, but forms covalent bonds
with the thymines if the sample is illuminated with UVA light.

Hoechst 33258 is a fluorescent stain that can bind to DNA by
intercalation or groove binding, with two different sets of
physicochemical parameters. In this case the drug concentration is
the factor that determines the dominant binding mode \cite{Silva}.
Some authors report a similar behavior for the intercalator
doxorubicin, have founding a possibility of groove binding at
AT-rich regions \cite{Arnaiz}.

\section{Connecting mechanics to physical chemistry}

In this section we introduce the main subject of this review, the
approaches developed to establish connections between the mechanical
properties and physicochemical properties of DNA-ligand complexes.
As stated before, the advantage in establishing this type of
connection is the possibility to deduce one or more properties of a
certain type (physicochemical properties, for example) knowing only
the behavior of a property of the other type (the persistence or
contour length, for example). With such connection(s), one can
considerably reduce the number of different experimental techniques
necessary to perform a robust characterization of the DNA
interaction(s) with a certain type of ligand. This fact thus reduces
the time and cost required for getting data, since less different
equipments are needed and the number of experiments that must be
conducted can be considerably reduced. Furthermore, and perhaps more
important, the approach opens the possibility of comparing data
obtained by means of very different experimental techniques,
increasing confidence in the results.

In single molecule stretching experiments performed by optical or
magnetic tweezers, the typical result obtained is the force $\times$
extension curve of the molecule, from where the mechanical
properties can be extracted by fitting an appropriate model (for
DNA, the WLC model). We will show that if one knows how the contour
and/or the persistence length varies as a function of the total
concentration of ligand in solution ($C_T$) (which is the amount of
ligand added in sample preparation), it is possible to deduce
physicochemical properties such as the equilibrium constants, the
cooperativity degree, the exclusion number, etc.

\subsection{Historical survey}

Along the past years many groups have used single molecule
techniques to identify the possible binding mechanisms of DNA-ligand
interactions and to extract physicochemical information of such
interactions from these types of experiments \cite{Dattagupta,
Fritzsche, Coury, Mihailovic, McCauley, McCauley2, Krautbauer,
Krautbauer2, Eckel, Tessmer, Sischka, Husale, Chaurasiya, Soler,
Allemand2, RochaJCP2, CrisafuliAPL, CrisafuliIB, Siman, Silva}.

To the best of our knowledge, the first attempt to connect mechanics
to physical chemistry for DNA-ligand systems was performed in the
early 80's by the group of D. M. Crothers, who have measured the
changes of the DNA contour length when interacting with various
drugs (netropsin, distamycin, iremycin, daunomycin) as a function of
the bound ligand fraction $r$, by using electric dichroism and a
phase partition technique \cite{Dattagupta, Fritzsche}.
Nevertheless, they have not directly determined physicochemical
properties from such data, a task which could only be performed with
complementary analyzes and/or techniques.

In 1996 Coury \emph{et al.} \cite{Coury} has determined
physicochemical properties from the contour length data of some
DNA-ligand complexes, obtained using AFM \cite{Coury}. In fact, by
determining the relative increase of the contour length of DNA
complexes formed with intercalators such as daunomycin and ethidium
bromide, the authors were capable to estimate binding parameters
such as the equilibrium constant and the exclusion number. Similar
approaches were used by Mihailovic \emph{et al.} \cite{Mihailovic}
and Rocha \emph{et al.} \cite{RochaJCP2} to extract physicochemical
information of DNA-intercalator complexes by measuring the relative
increase of the contour length, obtained using optical tweezers.

Basically, the idea to perform such task is the following. Call
$\Delta$ the natural distance between two DNA base pairs, which is
$\sim$ 0.34 nm for B-DNA. When an intercalating molecule binds to
this site, it increases such distance to a new value $\Delta$ +
$\delta$. Call $L_0$ the bare DNA contour length and $L$ the new
length for a certain amount of bound ligand represented by the bound
fraction $r$. One can promptly write the relation

\begin{equation}
L = L_0 + N_{b}\delta \label{incLintercal},
\end{equation}
where $N_b$ is the number of bound ligand molecules.

Observe that $L_0$ = $N_{bp}\Delta$, being $N_{bp}$ the number of
DNA base-pairs. Therefore one can write the relative change of the
contour length $\Theta$ as

\begin{equation}
\Theta = \frac{L - L_0}{L_0} = \frac{N_{b}\delta}{N_{bp}\Delta} =
\gamma r \label{incLintercal2},
\end{equation}
where $\gamma$ = $\delta$/$\Delta$.

With Eq. \ref{incLintercal2} one can directly connect the mechanical
parameter $L$ to physicochemical parameters by expressing the bound
ligand fraction $r$ by an adequate binding isotherm. In the case of
intercalators, the more convenient binding isotherm is the Neighbor
Exclusion Model (NEM) (see Section IIID), since this isotherm
captures in detail the neighbor exclusion effects which always
follow intercalative binding. Nevertheless, there are two problems
that must be bypassed to use this approach. The first one is that in
the NEM binding isotherm one cannot analytically isolate the
parameter $r$ to substitute in Eq. \ref{incLintercal2}. This
problem, however, can be bypassed with numerical approaches, as will
be discussed soon. The second and more serious problem is that, not
only NEM, but all binding isotherms are written as functions of the
free ligand concentration $C_f$, which is not a directly accessible
parameter. In fact, in general one knows only the total ligand
concentration in solution $C_T$, the quantity used to prepare the
sample, which is the sum of the bound and free ligand
concentrations, \textit{i. e.},

\begin{equation}
C_T = C_f + C_b \label{CtCfCb}.
\end{equation}

The partitioning of $C_T$ into $C_f$ and $C_b$ is not trivial to be
measured and one usually needs other experimental techniques
(microcalorimetry, absorption spectroscopy, equilibrium dialysis,
etc.) to evaluate such partitioning. There are, however, approaches
that can be performed to bypass this problem, allowing one to use
only single molecule stretching to characterize the interaction. In
fact, one can estimate the bound ligand concentration from contour
length changes if the length increase due to a single binding event
($\delta$) is known \cite{Coury, Mihailovic}. Alternatively, as a
first-order approximation one can consider $C_f$ $\sim$ $C_T$ in the
binding isotherm if the DNA concentration in the sample is very low
(because $C_b$ will also be very low in this case) \cite{Mihailovic,
Siman}. This approximation is much used in typical
tweezers experiments that tether an individual DNA molecule and
then rinse away any DNA molecules in solution prior to the
introduction of a ligand. Such approach is convenient because it
allows one to express the binding isotherm as a function of a
directly accessible parameter ($C_T$), although it cannot be used
always. A different approach was proposed originally by Rocha
\textit{et al.} in 2007 \cite{RochaJCP2}, which consists in
manipulating Eqs. \ref{CtCfCb}, \ref{incLintercal2} and
\ref{nemfinal} to write the relation

\begin{equation}
C_T = \frac{C_{bp}}{\gamma}\Theta + \frac{\Theta(\gamma - n\Theta +
\Theta)^{n-1}}{K_i(\gamma - n\Theta)^n} \label{ThetaRocha}.
\end{equation}

\begin{figure}
\centering
\includegraphics[width=6.5cm]{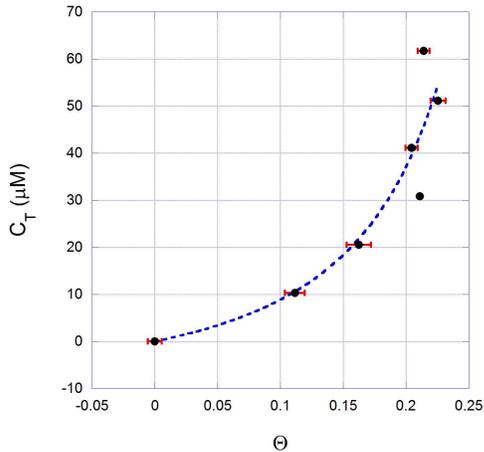}
\caption{Experimental result (\textit{circles}) of $C_T$ $\times$
$\Theta$ mesured for DNA complexes formed with the intercalator
diaminobenzidine, and a fitting to Eq. \ref{ThetaRocha}
(\textit{dashed line}). Observe that Eq. \ref{ThetaRocha} fits well
to the experimental data, returning the values of the
physicochemical parameters $N$ = 2.5 $\pm$ 0.6, $K_i$ = (1.8 $\pm$
0.6)$\times$10$^4$ M$^{-1}$ and $\gamma \sim$ 1. For this data
$C_{bp}$ = 2.4 $\mu$M.} \label{bindingdab}
\end{figure}

Such approach allows one to directly fit the contour length data
without any approximation: one should just plot the total ligand
concentration $C_T$ in the $y$-axis and the relative increase of the
contour length $\Theta$ in the $x$-axis, such that Eq.
\ref{ThetaRocha} can used directly to fit the experimental data. In
Fig. \ref{bindingdab} we show an example of such fitting, performed
originally in ref. \cite{Reis} for DNA complexes formed with the
intercalator diaminobenzidine. Other examples can be found in refs.
\cite{RochaJCP2} and \cite{RochaAPLPso} for DNA complexes formed
with the intercalators ethidium bromide and psoralen, respectively.

One should note, however, that the contour length approaches
discussed above can only be used for intercalators. In fact, only
intercalators increase the DNA contour length when binding
\cite{Sischka, Chaires, Fritzsche}. An exception are ligands that facilitate or inhibit base
pair formation, which can also be studied by
contour length approaches similar to those discussed above, using length changes as a marker of ligand binding
\cite{Chaurasiya2014}. The other common types of
interactions between DNA and ligands, such as groove binding,
electrostatic interaction or covalent binding in general do not
affect the DNA contour length. In some cases, however, these kinds
of interactions can cause DNA compaction with a decrease of the
``apparent contour length'' measured by force spectroscopy in the
low-force regime \cite{CrisafuliAPL, CrisafuliIB, Silva}. The
concept of ``apparent contour length'' arises from the fact that, if
the DNA molecule is partially compacted due to ligand binding, small
forces in the entropic regime usually are not sufficient to fully
stretch the molecule, and therefore the measured contour length will
be smaller than the real one. Depending on the type of interaction,
even high forces cannot be used to fully stretch the complexes and
estimate the real contour length by fitting the WLC model
\cite{Rocha2014}. The decrease of the ``apparent contour length''
upon increasing of the bound ligand concentration in general depends
on intricate effects such as the positional correlation of bound
ligands. This fact makes it difficult to directly link the contour length
data to a binding isotherm, although other kinds of analyses can be performed to study such interactions.  

On the other hand, the other basic mechanical property  (the persistence length)
is much more sensitive to other types of interactions, and
in general changes for covalent binding \cite{CrisafuliAPL,
CrisafuliIB, RochaAPLPso, Lee}, intercalative binding
\cite{RochaJCP2, Reis, Crisafuli2014, Sischka, Salerno, Cassina,
Tessmer} and groove/electrostatic binding \cite{Rouzina3, Manning,
Silva, Cesconetto, Sischka, McCauley, McCauley2}. This fact turns
the persistence length into the ideal mechanical property to be
choosen for monitoring DNA-ligand interactions and to be connected
with the physical chemistry of such interactions. Nevertheless, such
connection is not straightforward as the one performed for the
contour length of DNA-intercalator complexes.

Only in 1998 the first attempt to connect the persistence length to
physicochemical properties was performed by Rouzina and Bloomfield
\cite{Rouzina3}, which can be synthesized in Eq. \ref{modRouz}
presented earlier. This model however was derived in the context of
electrostatic interactions and attempt to explain the changes of the
persistence length due to the negative charge neutralization in the
DNA phosphate backbone \cite{Rouzina3}. Recently, this model has
been used to fit experimental data of DNA complexes formed with
positively charged proteins such as HMG, HMGB1 and HMGB2
\cite{McCauley, McCauley2}, with excellent agreement.

The question now is: Can the changes of the persistence length be
related to physicochemical parameters for any type of interaction?
In the next section we discuss an approach that can be used to
perform such task.

\subsection{A general model to connect the persistence length to physical chemistry}

The DNA molecule partially covered by ligand molecules along its
structure can be thought as an association of entropic springs in
series. One type of spring is the bare DNA with its natural
persistence length $A_0$, corresponding to the regions without bound
ligands along the contour length of the molecule. The other type(s)
of spring(s) is(are) the local complexe(s) formed between DNA and
the bound ligand molecules. A simple phenomenological model to study
the persistence length of DNA-ligand complexes that uses this
assumption was proposed by Rocha \cite{RochaPB}. Latter, it was
rigourously demonstrated \cite{Siman} that a series association of
$n$ entropic springs with persistence lengths $A_0$, $A_1$, $A_2$,
..., $A_{n - 1}$ results in an effective entropic spring with the
effective persistence length $A_E$ given by

\begin{equation}
\frac{1}{A_E} = \frac{f_0(r)}{A_0} + \frac{f_1(r)}{A_1} +
\frac{f_2(r)}{A_2} + ...\label{A_eff1},
\end{equation}
where $f_0$($r$), $f_1$($r$), $f_2$($r$), etc are specific functions
of the bound ligand fraction $r$.

The function $f_i$($r$) is in fact the probability of finding an
entropic spring (a part of the DNA molecule) along the contour
length with a local persistence length $A_i$ \cite{Siman}, which
depends on the bound site fraction $r$ \cite{McGhee, Siman}.

In general, the approach proposed in Eq. \ref{A_eff1} can be applied
by following three steps:

(a) One needs firstly to find the probability distribution of the
bound ligands, \textit{i. e.}, the set of functions $f_i$($r$).

(b) The second step is to choose an adequate binding isotherm that
captures the physical chemistry of the system, and then plug such
isotherm in Eq. \ref{A_eff1} via the parameter $r$.

(c) Finally, the third step is to use the equation constructed in
step (b) to fit the experimental data of the persistence length,
extracting the physicochemical parameters contained in the binding
isotherm and the set of local persistence lengths $A_i$'s.

To deduce the probability distribution mentioned in step (a), the
easiest way is firstly identify how many different entropic springs
one needs in the model to correct reproduce the experimental
behavior of the persistence length as a function of ligand
concentration. The simplest behavior of this parameter reported in
the literature is a monotonic decay, found for example for the
proteins HMG, HMGB1 and HMGB2 \cite{McCauley, McCauley2} and for the
drug cisplatin \cite{CrisafuliAPL, CrisafuliIB}. This relatively
simple behavior of the persistence length can be explained with a
model consisted only by two entropic springs, one representing the
bare DNA (local persistence length $A_0$) and the other representing
the local structure formed by the ligand molecule bound to the DNA
(local persistence length $A_1$).

From now on let us consider a ``site'' the place effectively
occupied by a single ligand molecule (or by a single bound cluster
of molecules, in the cases in which the ligands bind to DNA forming
clusters due to high positive cooperativity). One should note that
even for single-ligand binding the sites are usually higher than one
DNA base-pair, due to ligand size and/or neighbor-exclusion effects.
A model with only two different types of entropic springs, as
proposed in the last paragraph, is a \textit{one-site quenched
disorder statistical model}, since the probability distribution
depends only on the occupancy of \textit{single sites} along the
double-helix, \textit{i. e.}, it does not depend on the correlation
with the occupancy of nearest neighbor sites.

Consider now a particular site choose randomly along the DNA. The
probability of this site to be occupied by a ligand molecule is $x$
= $r$/$r_{max}$, with a local persistence length $A_1$; and the
probability of this site to be unoccupied is 1 - $x$, with a local
persistence length $A_0$ \cite{Siman, McGhee}. The effective
persistence length can then be written as

\begin{equation}
\frac{1}{A_E} = \frac{1 - x}{A_0} + \frac{x}{A_1} \label{A_eff2},
\end{equation}
and $x$ = $r$/$r_{max}$ can be directly connected to a binding
isotherm.

Equation \ref{A_eff2} was recently used by Crisafuli \textit{et al.}
to determine the physicochemical parameters of the DNA-cisplatin
interaction from the persistence length data of these complexes
\cite{CrisafuliAPL, CrisafuliIB}. One should observe that, since the
exclusion number $N$ is related to the saturated bound ligand
fraction $r_{max}$ by $N$ = 1/$r_{max}$, the electrostatic model
proposed by Rouzina and Bloomfield \cite{Rouzina3} (Eq.
\ref{modRouz}) is a particular case of Eq. \ref{A_eff2} (it is just
a matter of redefining the physical interpretation of the constants
$A_i$'s).

There are other types of ligands that can induce a more intricate
non-monotonic behavior for the persistence length as a function of
the ligand concentration. Probably the most known example is the
bacterial protein HU \cite{vanNoort}, but some drugs such as
catinonic cyclodextrins \cite{Siman}, actinomycin D
\cite{Cesconetto} and hoechst 33258 \cite{Silva} also induce such
behavior. To account for the persistence length changes of the DNA
complexes formed with these compounds, the one-site model discussed
above does not work, and one needs to introduce at least one more
entropic spring with other local persistence length ($A_2$),
\textit{i. e.}, one needs a \textit{two-sites quenched disorder
statistical model}, in which one must consider the probabilities
associated with the occupancy of two nearest sites. In the context
of a two sites model, there are therefore the following
probabilities associated to the local persistence lengths: (a) two
nearest sites unoccupied have local persistence length $A_0$ and
probability $P_0$ = (1-$x$)$^2$. (b) Two nearest sites simultaneous
occupied have local persistence length $A_2$ and probability $P_2$ =
$x$$^2$. (c) Finally, one site unoccupied and the neighbor occupied
have local persistence length $A_1$ and probability $P_1$ = 1 -
$P_0$ - $P_2$ = 2$x$(1-$x$). The effective persistence length can
therefore be written as

\begin{equation}
\frac{1}{A_E} = \frac{(1 - x)^2}{A_0} + \frac{2x(1-x)}{A_1} +
\frac{x^2}{A_2} \label{A_eff3},
\end{equation}
and $x$ = $r$/$r_{max}$ can be connected to a binding isotherm as
usual.

A last issue must be solved to complete the problem, both for
monotonic and non-monotonic behaviors of the persistence length: one
must write the binding isotherm as a function of a directly
accessible parameter instead of $C_f$, as discussed in Section VIA,
in order to eliminate the dependence in using other experimental
techniques to estimate the ligand partitioning between the DNA
($C_b$) and the solution ($C_f$). Although we have discussed some
approaches to perform this task for intercalators in Section VIA, it
is clear that a general approach is needed in order to contemplate
the order types of ligands.

In 2012 Siman \textit{et al.} have firstly proposed a simple
iterative solution of the binding isotherm \cite{Siman}, which was
promptly generalized by Cesconetto \textit{et al.} in 2013 with the
following method. Firstly choose a particular binding isotherm, for
example, the Hill binding isotherm (Eq. \ref{Hillfinal}). One can
plug the relations $x$ = $r$/$r_{max}$ and $C_f$ = $C_T$ - $rC_{bp}$
= $C_T$ - $r_{max}C_{bp}x$ in this binding isotherm to write

\begin{eqnarray}
x = \frac{[K_i(C_T - r_{max}C_{bp}x)]^n}{1+[K_i(C_T -
r_{max}C_{bp}x)]^n} \label{Hillx}.
\end{eqnarray}

Observe that this equation can be solved numerically for known
values of the constants, returning $x$ for each value of $C_T$.
Therefore, one needs to write a simple algorithm that uses a
subroutine to solve Eq. \ref{Hillx} for initial guessed values of
the constants, and uses the results returned for $x$ plugged into
Eq. \ref{A_eff3} or Eq. \ref{A_eff2} to fit the experimental data of
the persistence length $A$ as a function of $C_T$, by using least
squares fitting. With this approach the problem is completely
solved. Observe that any binding isotherm can be used to get an
equation similar to Eq. \ref{Hillx}, \textit{i. e.}, one needs only
to choose a plausible binding isotherm that captures the physical
chemistry of the interaction.

\begin{figure}
\centering
\includegraphics[width=6.5cm]{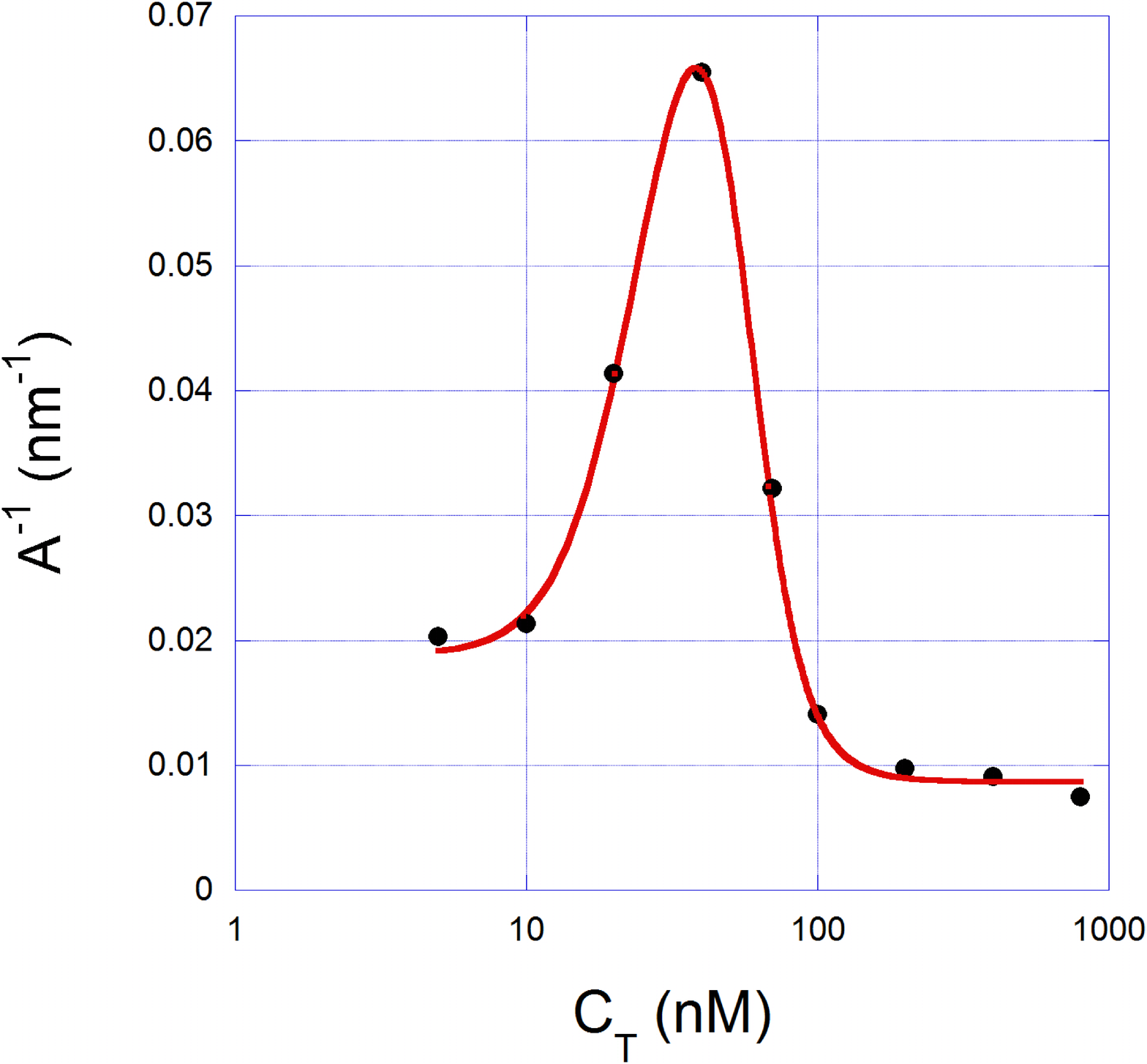}
\caption{\textit{Circles:} inverse of the persistence length of
DNA-HU complexes measured by single molecule stretching experiments
(experimental data by van Noort \textit{et al.} \cite{vanNoort},
error bars are not available in this case). \textit{Red solid line:}
a fitting to the model (Eq. \ref{A_eff3}) using the Hill binding
isotherm (Eq. \ref{Hillfinal}). From this fitting we have found the
physicochemical parameters $K_i$ = (3.4 $\pm$ 0.4) $\times$ 10$^7$ M
$^{-1}$, $n$ = 3.6 $\pm$ 0.3, $A_1$ = (8.5 $\pm$ 1) nm and $A_2$ =
(115 $\pm$ 13) nm. For this data $C_{bp}$ is unknown due to the
sample preparation procedure \cite{Siman}. It was left as an
adjustable parameter, and the fitting returns $C_{bp} \sim$ 110 nM.}
\label{HU}
\end{figure}

Below we revisit some results recently obtained with this approach,
showing that in principle it can be used to study any type of
interaction. The only requisite is that such interaction changes the
DNA persistence length as the ligand binds. All the experimental
data were obtained by single molecule stretching performed with
optical tweezers in the entropic regime, with fittings similar to that shown in Fig.
\ref{forceDNA} (except those of Fig. \ref{HU} - see ref.
\cite{vanNoort}). We also discuss the main features of the
physics and chemistry of the interactions revisited here. A complete
detailed discussion can be found in the original articles (and the
references therein). It is worth to emphasize that all
optical tweezers measurements were performed in chemical equilibrium, waiting
sufficient time for ligand equilibration before performing the
stretching experiments. In addition, these measurements were performed
with low forces ($<$ 2 pN) and pulling rates ($\sim$ 0.1 $\mu$m/s) in order to guarantee
that the chemical equilibrium is not affected by the stretching forces.

\begin{figure}
\centering
\includegraphics[width=6.5cm]{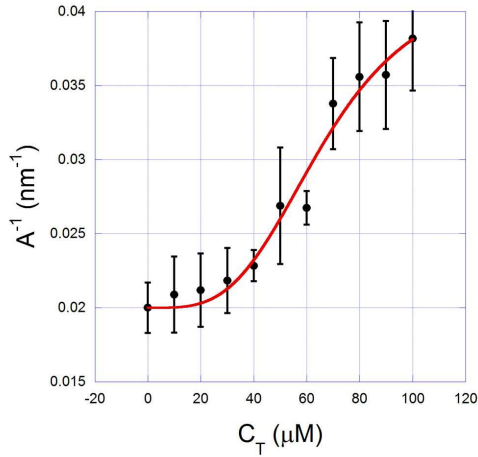}
\caption{\textit{Circles:} inverse of the persistence length of
DNA-cisplatin complexes measured by single molecule stretching
experiments. \textit{Red solid line:} a fitting to the model (Eq.
\ref{A_eff2}) using the Hill binding isotherm (Eq. \ref{Hillfinal}).
From this fitting we have found the physicochemical parameters $K_i$
= (1.6 $\pm$ 0.2) $\times$ 10$^4$ M $^{-1}$, $n$ = 3.6 $\pm$ 0.4,
$r_{max}$ = 0.56 $\pm$ 0.06 and $A_1$ = (24 $\pm$ 4) nm. For this
data $C_{bp}$ = 8.9 $\mu$M.} \label{cisplatin}
\end{figure}

In Fig. \ref{cisplatin} we show the experimental data
(\textit{circles}) of the persistence length of DNA-cisplatin
complexes as a function of drug total concentration in the sample
$C_T$. Observe that, for convenience to fit with Eq. \ref{A_eff2},
we have plotted the inverse of the persistence length in this figure
and in all subsequent ones. The fitting with the model (Eq.
\ref{A_eff2}) is also shown (\textit{red solid line}). In this case
we have used the Hill binding isotherm (Eq. \ref{Hillfinal}) to
perform the fitting, extracting the physicochemical parameters $K_i$
= (1.6 $\pm$ 0.2) $\times$ 10$^4$ M $^{-1}$, $n$ = 3.6 $\pm$ 0.4,
$r_{max}$ = 0.56 $\pm$ 0.06 and $A_1$ = (24 $\pm$ 4) nm. These
results agree very well to those presented in ref.
\cite{CrisafuliIB}, which were obtained using another fitting
strategy, and as well as to results obtained from other experimental
techniques \cite{Takahara, Zamble}. In particular, the Hill exponent
$n$ = 3.6 indicates that cisplatin presents positive cooperativity
in its interaction with DNA.

Cisplatin and its analogues carboplatin and oxaliplatin form one of
the most important class os compounds used in cancer chemotherapies,
especially to treat head, neck, testicular, ovarian and non-small
cell lung cancers \cite{Chaney}. In aqueous solution, two chloride
ions dissociate from the compound, followed by incorporation of two
water molecules. This is the active state of the drug, which can
bind to DNA \cite{Pinto}. Many aspects of the DNA-cisplatin
interaction are currently well established in the literature, such
as the mechanism of action of the compound as an anticancer drug,
which consists in damaging the DNA molecule with adducts that form
interstrand and intrastrand crosslinks \cite{Stehlikova}. These
crosslinks hinder DNA replication by introducing strong
structural perturbations on the double-helix such as bendings,
partial unwinding and loops \cite{Stehlikova, Hou, Lee}. These
structural perturbations are closely related to the result found for
the Hill exponent ($n \sim$ 3.6). In fact, a positive cooperativity
could be expected in DNA-cisplatin interaction, since the crosslinks
and loops induced in the DNA by the drug approximate different
strand segments as the drug concentration is increased, therefore
increasing the probability of forming even more crosslinks and loops
as cisplatin binds \cite{CrisafuliAPL, CrisafuliIB}. A nearly
similar mechanism was recently observed for the H-NS binding protein
by Dame \emph{et al.}, which have shown that a cooperative behavior
in this case arises as an intrinsic property of DNA bridging due to
duplex proximity \cite{Dame}.

\begin{figure}
\centering
\includegraphics[width=6.5cm]{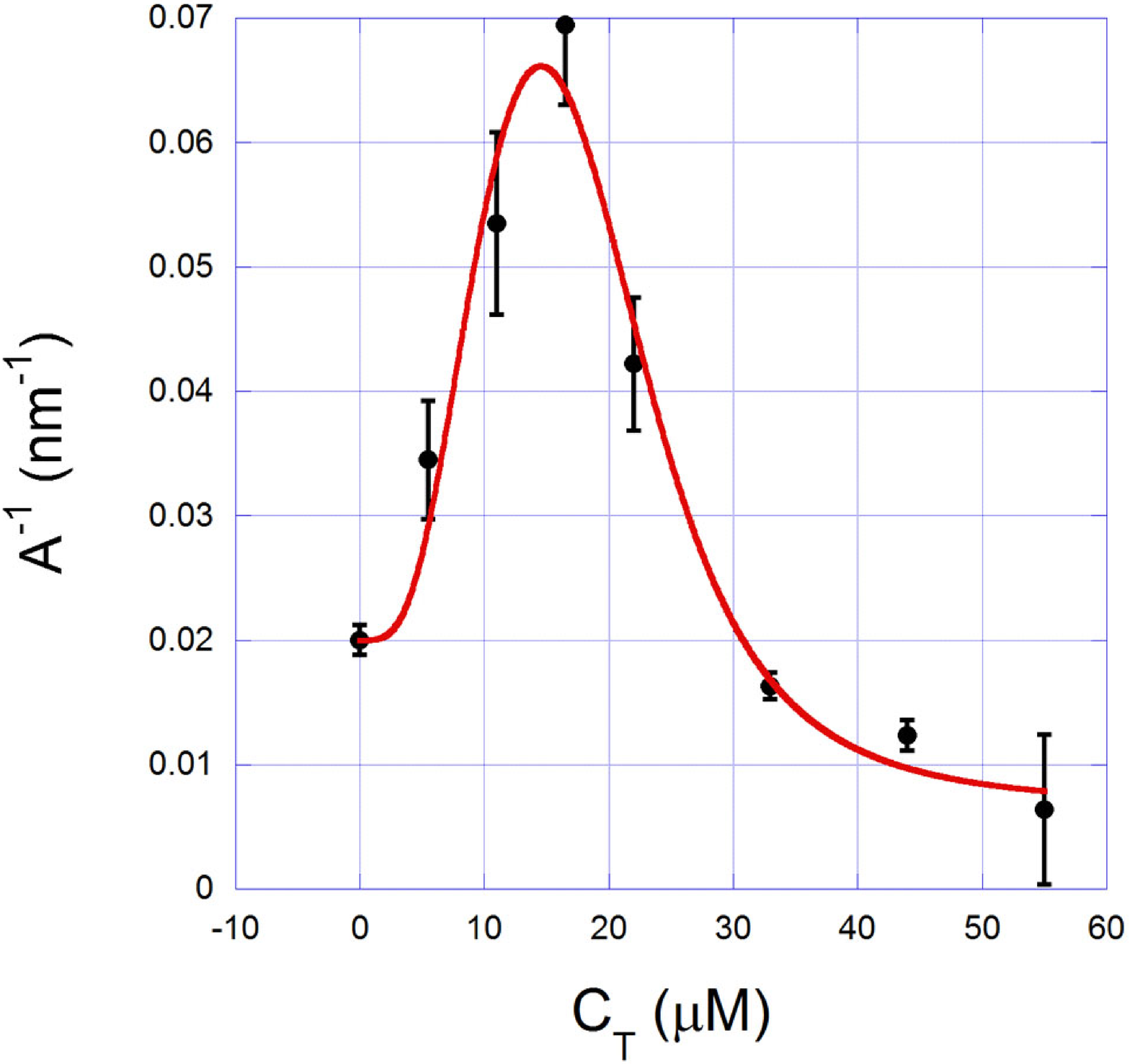}
\caption{\textit{Circles:} inverse of the persistence length of
DNA-cyclodextrin complexes measured by single molecule stretching
experiments. \textit{Red solid line:} a fitting to the model (Eq.
\ref{A_eff3}) using the Hill binding isotherm (Eq. \ref{Hillfinal}).
From this fitting we have found the physicochemical parameters $K_i$
= (9 $\pm$ 1) $\times$ 10$^4$ M $^{-1}$, $n$ = 3.7 $\pm$ 0.4,  $A_1$
= 8.4 $\pm$ 1 nm and $A_2$ = 149 $\pm$ 21 nm. For this data $C_{bp}$
= 11 $\mu$M.} \label{cyclodextrin}
\end{figure}

In Fig. \ref{cyclodextrin} we show the experimental data
(\textit{circles}) for the inverse of the persistence length of DNA
complexes formed with a monovalent cationic $\beta$-cyclodextrin
(6-monodeoxy-6-monoamine-$\beta$-cyclodextrin) as a function of drug
total concentration in the sample $C_T$, firstly presented in ref.
\cite{Siman}. Cyclodextrins (CDs) are cyclic oligosaccharides
composed of D-glucose units joined by glucosidic linkages. The
$\beta$ subtype consists of seven units and has a structure that
resembles a truncated cone, with hydroxyl groups localized at the
outer surface of the cone. That gives CDs the property to be
water-soluble and to have a relatively hydrophobic inner cavity able
to partially or entirely accommodate polymers forming host-guest
inclusion complexes \cite{Spies}. Monovalent cationic
$\beta$-cyclodextrin is usually obtained by substituting one of the
hydroxyl groups by an amino group. This molecule has been used for
condensing DNA and introducing it into small vesicles for gene
therapy applications \cite{Tavares}.

Observe in Fig. \ref{cyclodextrin} that for the complexes formed
between DNA and cationic $\beta$-cyclodextrin the persistence length
exhibits a non-monotonic behavior, and therefore we have used Eq.
\ref{A_eff3} to fit the data, together with the Hill binding
isotherm (\textit{red solid line}). We have found the results $K_i$
= (9 $\pm$ 1) $\times$ 10$^4$ M $^{-1}$, $n$ = 3.7 $\pm$ 0.4,  $A_1$
= (8.4 $\pm$ 1) nm and $A_2$ = (149 $\pm$ 21) nm. The parameter
$r_{max}$ = 0.67 was known for this ligand such that we have fixed
its value in the fitting \cite{Siman}. The value obtained for the
Hill exponent $n$ again indicates that the system is positively
cooperative, in this case forming bound clusters of $\sim$ 4 drug
molecules at the binding sites \cite{Siman}.

The results obtained for the bacterial protein HU in ref.
\cite{Siman} are somewhat similar, as shown in Fig. \ref{HU}. We
have used again Eq. \ref{A_eff3} and the Hill binding isotherm to
perform the fitting (\textit{red solid line}), and the experimental
data (\textit{circles}) were obtained by van Noort \textit{et al.}
for this ligand \cite{vanNoort}. From the fitting we have obtained
the results $K_i$ = (3.4 $\pm$ 0.4) $\times$ 10$^7$ M $^{-1}$, $n$ =
3.6 $\pm$ 0.3, $A_1$ = (8.5 $\pm$ 1) nm and $A_2$ = (115 $\pm$ 13)
nm. Here again, $r_{max}$ = 0.11 is a known parameter and was
maintained fixed in the fitting \cite{Siman}. For this data $C_{bp}$
is unknown due to the sample preparation procedure \cite{Siman}. It
was left as an adjustable parameter, and the fitting returns $C_{bp}
\sim$ 110 nM. Here the fact that the persistence length increases
for high protein concentrations agrees with results obtained in AFM
images, which have shown the formation of rigid filaments
\cite{vanNoort}.

At this point it is necessary to reflect on the use of the Hill
binding isotherm in the analysis of DNA-ligand systems. The fact
that we have found a Hill exponent $n \gg$ 1 for cisplatin,
cyclodextrin and HU strongly indicates that relevant positive
cooperativity is present in such systems. In fact, binding isotherms
with no cooperativity such as the Scatchard model (Eq.
\ref{Scatfinal}) or the basic Neighbor Exclusion Model (NEM) (Eq.
\ref{nemfinal}) do not work in performing these fittings. The
cooperative version of the neighbor exclusion model (Eq.
\ref{nemcoopfin}) in principle could be used, but we were not
successful in performing the fitting in an easy way, founding
numerical problems in solving the equation analogue to Eq.
\ref{Hillx} for this binding isotherm. In fact, the intricacy of Eq.
\ref{nemcoopfin} somewhat limits its applicability in the fitting
approaches discussed in this review. For this reason we use the Hill
binding isotherm, a much simpler equation that also takes into
account cooperativity effects.

The first example of a non-cooperative system studied with our
approach are the DNA complexes formed with the drug Actinomycin D
(ActD), firstly presented in ref. \cite{Cesconetto}. This drug is a
DNA ligand clinically used as an antibiotic and to treat some highly
malignant cancers, such as gestational trophoblastic disease
\cite{Osathanondh}, Wilms' tumor \cite{DAngio} and rhabdomyosarcoma
\cite{Pinkel}. The drug exhibits a complex interaction with
double-strand DNA, presenting two distinct parts which bind to DNA
by different modes: while the phenoxazone ring intercalates,
preferentially at the CG base pairs, the cyclic pentapeptide chains
bind to the minor groove, usually forming hydrogen bonds with the
guanine bases \cite{Patel, Muller, Sobell, Takusagawa}.

\begin{figure}
\centering
\includegraphics[width=6.5cm]{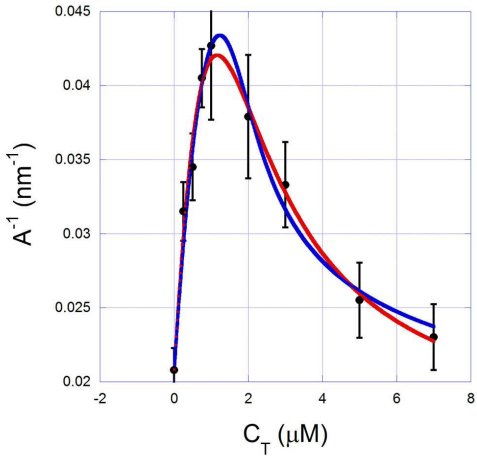}
\caption{\textit{Circles:} inverse of the persistence length of
DNA-ActD complexes measured by single molecule stretching
experiments. \textit{Red solid line:} a fitting to the model (Eq.
\ref{A_eff3}) using the Hill binding isotherm (Eq. \ref{Hillfinal}).
From this fitting we have found the physicochemical parameters $K_i$
= (1.5 $\pm$ 0.4) $\times$ 10$^6$ M $^{-1}$, $n$ = 1.1 $\pm$ 0.2,
$r_{max}$ = 0.11 $\pm$ 0.01, $A_1$ = (15.2 $\pm$ 0.6) nm and $A_2$ =
(64 $\pm$ 25) nm. \textit{Blue solid line:} a fitting to the model
(Eq. \ref{A_eff3}) using the NEM binding isotherm (Eq.
\ref{nemfinal}). From this fitting we have found the physicochemical
parameters $K_i$ = (4.6 $\pm$ 0.5) $\times$ 10$^6$ M $^{-1}$, $N$ =
4 $\pm$ 0.5 (the exclusion number for each bound ActD), $A_1$ = (14
$\pm$ 2) nm and $A_2$ = (140 $\pm$ 16) nm. For this data $C_{bp}$ =
10.6 $\mu$M.} \label{ActD_A}
\end{figure}

Here we clearly have the option of choosing different binding
isotherms to perform the fitting. This fact illustrates the
versatility of our approach, which returns consistent results even
for different binding isotherms: it is required only to choose one
that captures the basic physical chemistry of the system. In fact,
if the DNA-ActD interaction is non-cooperative \cite{Cesconetto},
one can choose the Scatchard model or the basic (non-cooperative)
neighbor exclusion model. Nevertheless, instead of the first option
(Scatchard), we have chosen the Hill binding isotherm to perform the
fitting. If everything is right, one should find a Hill exponent
near unity ($n \sim$ 1), since in this case the Hill model is just
equivalent to the Scatchard one. Figure \ref{ActD_A} shows the
experimental data points (\textit{circles}) and the fittings to Eq.
\ref{A_eff3} with the Hill model (\textit{red solid line}) and with
the neighbor exclusion model (\textit{blue solid line}). From the
first fitting (Hill), we find $K_i$ = (1.5 $\pm$ 0.4) $\times$
10$^6$ M $^{-1}$, $n$ = 1.1 $\pm$ 0.2, $r_{max}$ = 0.11 $\pm$ 0.01,
$A_1$ = (15.2 $\pm$ 0.6) nm and $A_2$ = (64 $\pm$ 25) nm. From the
second fitting (NEM) we find $K_i$ = (4.6 $\pm$ 0.5) $\times$ 10$^6$
M $^{-1}$, $N$ = 4 $\pm$ 0.5 (the exclusion number for each bound
ActD), $A_1$ = (14 $\pm$ 2) nm and $A_2$ = (140 $\pm$ 16) nm.
Observe that both fittings explain well the behavior of the
experimental data. The results returned for the physicochemical
parameters, although somewhat dependent on the chosen binding
isotherm, are realist. The relatively high variability on the values
found for some of these parameters is compatible to the variability
found when using different experimental techniques
\cite{Paramanathan, Ruggiero, Sha, Goodisman}.

A relevant question that can be raised at this point is about the
accuracy of our approach to treat systems with more than one binding
mode, \textit{i. e.}, with two or more different sets of
physicochemical parameters. Many compounds interact with DNA in this
way, and recently we have successfully applied our fitting approach
to the fluorescent dye Hoechst 33258 \cite{Silva}.

The Hoechst stains, also known as bis-benzimides, are a family of
fluorescent dyes largely  employed to stain the DNA molecule in
molecular biology applications, allowing one to visualize DNA
with fluorescence microscopy. In addition, these compounds can be
potentially used as anticancer drugs \cite{Reddy}, since their
strong interaction with DNA can impede the replication of the
molecule. Many experimental techniques were employed over the past
years to study the effects of the Hoechst 33258 subtype on the DNA
molecule. In particular, it was found that the ligand binds preferentially
to the DNA minor groove, especially at AT-rich regions \cite{Saito,
Bailly}. Nevertheless, some authors have proposed that the ligand
presents more than one binding mode to double-strand (ds) DNA
\cite{Bontemps, Stokke, Bailly}, indicating the possibility of
intercalation at GC-rich regions \cite{Colson, Bailly}.

\begin{figure}
\centering
\includegraphics[width=6.5cm]{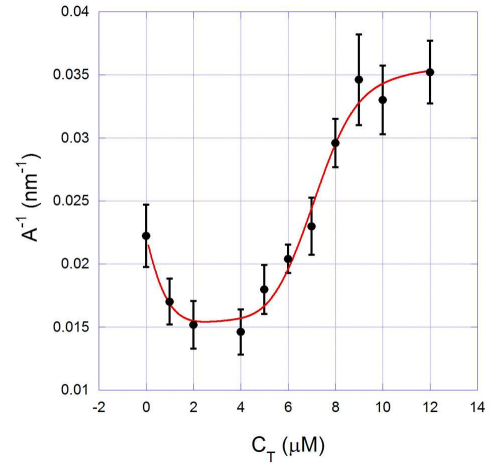}
\caption{\textit{Circles:} inverse of the persistence length of
DNA-hoechst complexes measured by single molecule stretching
experiments. \textit{Red solid line:} a fitting to the model (Eq.
\ref{A_eff3}) using a sum of two Hill processes as the binding
isotherm. From this fitting we decouple the two binding modes and
find the physicochemical parameters  $K_1$ = (1.8 $\pm$ 0.4)
$\times$ 10$^6$ M $^{-1}$, $n_1$ = 1.1 $\pm$ 0.3, $K_2$ = (2.4 $\pm$
0.2) $\times$ 10$^5$ M $^{-1}$, $n_2$ = 7 $\pm$ 3, $A_1$ = (300
$\pm$ 100) nm and $r_{max}$ = 0.32 $\pm$ 0.02. For this data
$C_{bp}$ = 20 $\mu$M.} \label{hoechst}
\end{figure}

With our fitting approach we were able to decouple the two main
binding modes that Hoechst 33258 exhibits with DNA, by using a
binding isotherm expressed as a sum of two Hill processes. We have
determined the two complete sets of physicochemical parameters for
each of the binding modes. In particular, we have found that the
first binding mode (intercalation) is non-cooperative, with a Hill
exponent $\sim$ 1, while the second mode (groove binding) is highly
positively cooperative, with a Hill exponent $\sim$ 7. Such
conclusion is in agreement with previous studies performed by other
techniques (equilibrium dialysis and absorption spectroscopy)
\cite{Bontemps, Stokke}. The two binding modes coexist in the entire
concentration range studied here, but intercalation is dominant for
$C_T <$ 3 $\mu$M while groove binding is dominant for higher
concentrations. Figure \ref{hoechst} shows the experimental data
(\textit{circles}) and the fitting (\textit{red solid line}). We
have found that, for the intercalative binding mode, $K_1$ = (1.8
$\pm$ 0.4) $\times$ 10$^6$ M $^{-1}$, $n_1$ = 1.1 $\pm$ 0.3. On the
other hand, for minor groove binding, we found $K_2$ = (2.4 $\pm$
0.2) $\times$ 10$^5$ M $^{-1}$, $n_2$ = 7 $\pm$ 3. Also, we found
$A_1$ = (300 $\pm$ 100) nm and $r_{max}$ = 0.32 $\pm$ 0.02, which
are global parameters independent of the binding mode. The parameter
$A_2$ = 28.3 nm was maintained fixed in the fitting since it is the
saturation value of the persistence length, which can be directly
determined from the data of Fig. \ref{hoechst} in this case. The
error bars of the parameters obtained in this fitting are somewhat
higher than those obtained for the other DNA-ligand systems
presented above. This fact is due to the excess of adjustable
parameters used in the fitting procedure in this case, because we
have two different binding modes and consequently two sets of
binding parameters.

Finally we show an example of how our fitting approach can also be
used to analyze the contour length data of DNA complexes formed with
ligands. GelRed is a fluorescent nucleic acid stain designed with
the purpose of replacing the highly toxic ethidium bromide (EtBr) in
gel electrophoresis and other experimental techniques which depends
on the fluorescence of stained DNA. When bound to DNA, GelRed has
the same absorption and emission spectra of EtBr and, according to
its manufacturer (Biotium Inc., Hayward, CA, USA), it has the
advantage of being  much less toxic and mutagenic \cite{GRSafety,
Huang}.

\begin{figure}
\centering
\includegraphics[width=6.5cm]{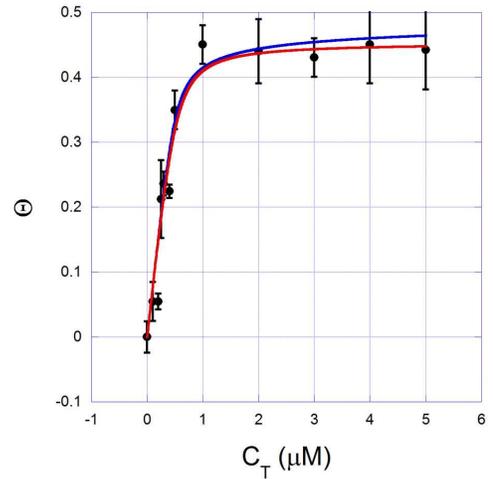}
\caption{\textit{Circles:} experimental data of the relative
increase of the contour length $\Theta$ = ($L$ - $L_0$)/$L_0$ of
DNA-GelRed complexes. \textit{Red solid line:} a fitting to Eq.
\ref{incLintercal2} using the Scatchard binding isotherm (Eq.
\ref{Scatfinalrmax}), from which we have found the physicochemical
parameters $K_i$ = (1.8 $\pm$ 0.4) $\times$ 10$^7$ M $^{-1}$,
$r_{max}$ = 0.22 $\pm$ 0.03 and $\gamma$ = 2.2 $\pm$ 0.1.
\textit{Blue solid line:} a fitting to Eq. \ref{incLintercal2} using
the NEM binding isotherm (Eq. \ref{nemfinal}), from which we have
found the physicochemical parameters $K_i$ = (1.8 $\pm$ 0.3)
$\times$ 10$^7$ M $^{-1}$, $N$ = 3.7 $\pm$ 0.4 and $\gamma$ = 1.9
$\pm$ 0.1. For this data $C_{bp}$ = 2.4 $\mu$M.} \label{GelRed}
\end{figure}

Figure \ref{GelRed} shows the experimental data of the relative
increase of the contour length $\Theta$ = ($L$ - $L_0$)/$L_0$
(\textit{circles}), firstly presented in ref. \cite{Crisafuli2014},
and two fittings performed with Eq. \ref{incLintercal2} and two
different binding isotherms: Scatchard (\textit{red solid line}) and
NEM (\textit{blue solid line}). The two fittings are similar,
returning equivalent physicochemical parameters and allowing one to
conclude that GelRed interacts with DNA by bis-intercalation
\cite{Crisafuli2014}. For the Scatchard fitting, we have found $K_i$
= (1.8 $\pm$ 0.4) $\times$ 10$^7$ M $^{-1}$, $r_{max}$ = 0.22 $\pm$
0.03 and $\gamma$ = 2.2 $\pm$ 0.1. For the NEM fitting, we found
$K_i$ = (1.8 $\pm$ 0.3) $\times$ 10$^7$ M $^{-1}$, $N$ = 3.7 $\pm$
0.4 and $\gamma$ = 1.9 $\pm$ 0.1. The results obtained for this
ligand with our fitting approach lead us to conclude that the GelRed
dye is a bis-intercalator. In fact, the exclusion parameter $N$ =
1/$r_{max}$ indicates that each bound GelRed molecule effectively
occupies $\sim$ 4 DNA base-pairs, a value considerably higher than
the results found for most monointercalators, and approximately
twice the result for EtBr (which is $\sim$ 2 \cite{RochaJCP2,
Chaires, Gaugain}). The equilibrium association constant $K_i$ is
also higher than the result obtained for typical monointercalators
($\sim$ 10$^5$ M$^{-1}$) \cite{RochaJCP2, RochaPB, Chaires,
Gaugain}, and within the range found for most bis-intercalators
(10$^7$ to 10$^9$ M$^{-1}$) \cite{Gunther, Berge, Murade, Maaloum,
Garbay}. Finally, the result $\gamma \sim$ 2  is approximately twice
the value obtained for typical monointercalators, suggesting that
each bound GelRed molecule increases the DNA contour length by
$\sim$ 0.68 nm, a result also compatible to typical
bis-intercalators \cite{Gunther, Maaloum}. Observe that the
bis-intercalators should increase approximately twice the DNA
contour length per bound molecule, since each ligand molecule
contains two intercalating portions.

In summary, we have presented many examples of DNA-ligand systems
analyzed with the proposed fitting approach. All the results
obtained for the physicochemical parameters are consistent with most
studies found in the literature that have used many different
experimental techniques, from crystallography to fluorescence
resonance energy transfer \cite{Takahara, Sagi, Rappaport,
Paramanathan, Galo, Ruggiero, Sha, Goodisman, Bontemps, Stokke}.
Thus, our fitting approach allows a direct comparison between the
results obtained from single molecule stretching experiments to
those obtained from typical ensemble-averaging techniques, which are
usually used to characterize the physical chemistry of DNA-ligand
interactions. A weakness of the presented approach that can be
pointed concerns the error bars of the physicochemical parameters
obtained from the fitting procedure, which can be a bit high if the
number of adjustable parameters used in the fitting is high, as in
the case of multiple binding modes. Nevertheless in these cases one
can maintain fixed some parameters previously measured with other
techniques and perform the fitting using only the adjustable
parameters of interest. In this way, the fitting approach can be
used to compare and verify results obtained from very different
experimental techniques, and can still be useful in the
investigation of DNA-ligand interactions.

\section{Conclusions}

We have reviewed important topics of the field ``DNA-ligand
interactions'', from DNA mechanics to DNA-ligand physical chemistry,
emphasizing how one can connect the changes of the mechanical
properties of DNA induced by the binding ligand to the
physicochemical information of such interaction. This type of
connection is extremely relevant because it allows one to perform a
robust characterization of the interaction both from the point of
view of the mechanical properties and of the physical chemistry of
the interaction by using only one experimental technique: single
molecule stretching experiments. Moreover, the possibility of
performing such connection reduces the time and cost required for
getting results about a DNA-ligand system, since less different
equipments are required and the number of experiments that must be
conducted can be considerably reduced. Furthermore, and more
important, it opens the possibility of comparing the results
obtained by means of very different experimental techniques, in
special when comparing single molecule techniques to
ensemble-averaging techniques.

In particular, we reviewed a fitting approach recently proposed by
our group to connect the persistence length of the DNA-ligand
complexes to the physical chemistry of the interaction. Such
approach in principle can be used for any type of ligand, from drugs
to proteins, even if there are multiple binding modes. However, a
test with sequence-specific ligands \cite{Waring} is still needed.
In any case, the only requisite to try the approach is that the
interaction must change the DNA persistence length as the ligand
binds, which usually occurs for all types of common interactions
(intercalation, covalent binding, electrostatic driven interactions
and groove binding) at least for some ligand concentration range.

\section{Acknowledgements}

This work was supported by the Brazilian agencies: Funda\c{c}\~ao de
Amparo \`a Pesquisa do Estado de Minas Gerais (FAPEMIG), Conselho
Nacional de Desenvolvimento Cient\'ifico e Tecnol\'ogico (CNPq) and
Coordena\c{c}\~ao de Aperfei\c{c}oamento de Pessoal de N\'ivel
Superior (CAPES). The author also thanks the collaborators and
students that have contributed to many of the presented results: E.
B. Ramos, O. N. Mesquita, F. A. P. Crisafuli, E. F. Silva, E. C.
Cesconetto and L. Siman.

\bibliography{Rocha2015_bibtex}

\end{document}